\documentclass[%
 reprint,
amsmath,amssymb,
aps,
prb,
5prl
floatfix
]{revtex4-1}

\usepackage{graphicx}
\usepackage{dcolumn}
\usepackage{bm}

\usepackage[colorlinks=true,linkcolor=Red,citecolor=Green,linktoc=page,hypertex]{hyperref}
\usepackage[usenames,dvipsnames]{color}

\usepackage{caption}
\usepackage{float}

\begin{document}

\preprint{APS/123-QED}

\title{Finite quasiparticle lifetime in disordered superconductors}

\begin{abstract}

We investigate the complex conductivity of a highly disordered
MoC superconducting film with $k_Fl\approx 1$,
where $k_F$ is the Fermi wavenumber and $l$ is the mean free path,
derived from experimental transmission characteristics of coplanar waveguide resonators
in a wide temperature range below the superconducting transition temperature $T_c$.
We find that the original Mattis-Bardeen model with a finite quasiparticle lifetime,
$\tau$, offers a perfect description of the experimentally observed complex conductivity.
We show that $\tau$ is  appreciably reduced by scattering effects.
Characteristics of the scattering centers are independently found
by the scanning tunneling spectroscopy and agree with those determined from the
complex conductivity.
\end{abstract}

\date{\today}
\author{M.~\v{Z}emli\v{c}ka, P.~Neilinger, M.~Trgala, M. Reh\'{a}k, D.~Manca, M.~Grajcar}
\affiliation{Department of Experimental Physics, Comenius University, SK-84248 Bratislava, Slovakia}
\affiliation{Institute of Physics, Slovak Academy of Sciences, D\'{u}bravsk\'{a} cesta, Bratislava, Slovakia}
\author{P.~Szab\'{o}, P.~Samuely}
\affiliation{Institute of Experimental Physics, Slovak Academy of Sciences, $\&$ FS UPJ\v{S} , Watsonova 47, SK-04001 Ko\v{s}ice, Slovakia}
\author{{\v{S}.~Ga\v{z}}i}
\affiliation{Institute of Electrical Engineering, Slovak Academy of Sciences
,D\'{u}bravsk\'{a} cesta, SK-84104 Bratislava, Slovakia}
\author{U. H\"{u}bner}
\affiliation{Leibniz Institute of Photonic Technology, P.O. Box 100239, D-07702 Jena, Germany}
\author{V.\,M.\,Vinokur}
\affiliation{Argonne National Laboratory, 9700 S. Cass Ave, Argonne, IL 60439, USA }
\author{E. Il'ichev}
\affiliation{Leibniz Institute of Photonic Technology, P.O. Box 100239, D-07702 Jena, Germany}
\affiliation{Novosibirsk State Technical University, 20 Karl Marx Avenue, 630092 Novosibirsk, Russia}
\pacs{\ldots}
\maketitle

\section{INTRODUCTION}
Disordered superconductors are a subject of intense current attention.
This interest is motivated not only by the appeal of dealing with the most
fundamental issues of condensed matter physics involving interplay of quantum correlations,
disorder, quantum and thermal fluctuations, and Coulomb interactions,\cite{Escoffier04,Vinokur08,Sacepe08,Zaikin97} but also by the high promise for applications.
The existence of states with giant capacitance and inductance in the critical vicinity
of superconductor-insulator transition\cite{Vinokur08,Astafiev12} breaks ground for
novel microwave engineering exploring duality between phase slips at
point-like centers\cite{Tinkham} or at phase slip lines\cite{Ilichev92} and Cooper pair tunneling.\cite{Mooij06,Baturina13a,Baturina13b}
The feasibility of building a superconducting flux qubit by employing quantum phase slips in a weak link created by highly disordered superconducting wire was demonstrated by Astafiev \textit{et al}\cite{Astafiev12}.
Yet, while there has been notable recent success in describing DC properties of disordered superconductors,\cite{Goldman:2010,Vinokur08,Baturina13a}
the understanding of their AC response remains insufficient and impedes advance in their microwave applications.

Recent studies of the electromagnetic response of strongly disordered superconducting films\cite{Driessen12,Coumou13} revealed a discrepancy between the 
local density of states measured by scanning tunneling spectroscopy and the density of states that
has been assumed to describe the microwave response. This implies that a model assuming uniform properties of the film fails 
to describe films near Ioffe-Regel limit, $k_Fl\approx 1$. And
although the authors succeeded to explain the behavior of the imaginary part of the complex conductivity $\sigma_s=\sigma_1-i\sigma_2$ in a narrow temperature range, the understanding of the real part 
$\sigma_1$, which is mostly influenced by disorder, is far from being complete.
Hence a call arises for a simple unified model, that could explain both the microwave and
the tunneling conductance measurements in strongly disordered superconducting films.
In this paper, we discuss a model that meets this challenge and experimentally demonstrate its validity.

\section{MoC thin film and CPW resonator}
\begin{figure}
	\centering
	\includegraphics[width=8cm]{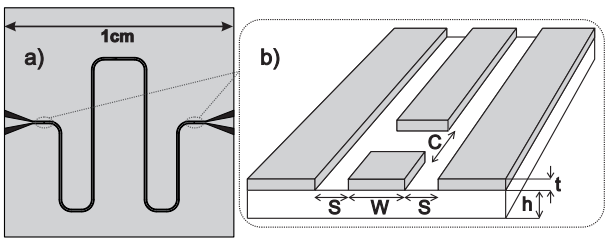}
	\caption{Scheme of the coplanar waveguide resonator with dimensions: W=50~$\mu$m, S=30~$\mu$m, C=10~$\mu$m, t=10~nm, h=430~$\mu$m. \label{fig:cpw}}
\end{figure}

One of the ways to measure the microwave complex conductivity is to use a coplanar waveguide (CPW) resonator patterned on a thin film of desired superconductor. The resonator is characterized by two main quantities, the resonant frequency and the quality factor, which can be directly calculated from the complex conductivity.
The capacitance of the CPW is explicitly defined by its geometry.\cite{goppl08} The imaginary part of the impedance is mostly represented by the inductance of the CPW and therefore, it determines CPW resonator resonant frequency. The real part of the impedance is determined by the resistive losses in the CPW and therefore influences the internal quality factor.\cite{goppl08} Taking into account the external quality factor $Q_{ext}$ due to input/output coupling capacitances, one can calculate the required  loaded quality factor. The structure of the CPW resonator (see Fig.\,\ref{fig:cpw}) was patterned by optical lithography and argon ion etching of the deposited superconducting  thin films. We focused on the study of the properties of disordered  10\,nm thin MoC films with sheet resistance  $R_{\Box}\approx 180\ \Omega$.\cite{Lee90} The chosen thickness of 10~nm is optimal for further patterning of superconducting nanostructures which are expected to exhibit quantum phase slips.\cite{Astafiev12}
The films were fabricated by magnetron reactive sputtering, where particles of molybdenum were sputtered from a Mo target onto sapphire $c$-cut substrate in argon-acetylene atmosphere. The partial pressure of acetylene and Ar gas was set to $3\times 10^{-4}$~mbar and $5.4\times 10^{-3}$~mbar, respectively.
The film thickness was controlled by tuning the sputtering time according to the sputtering rate of 10~nm/min. The RMS (root mean square) roughness of the surface 
1~$\mu$m$~\times~$1~$\mu$m calculated from the AFM topography data was about 0.3 nm. The preparation details are given in Ref.~\onlinecite{Trgala13}.

The transport properties of the MoC thin films have been obtained by four-probe measurements. The critical temperatures of superconducting transition $T_c$, are very sharp, showing a shift from 8~K for film thickness $t \ge 30$~nm down to 1.3~K for $t=3$~nm accompanied by an increase of the sheet resistance from several tens of Ohms to 1300~Ohms, respectively. The transport  measurements in magnetic fields as well as the Hall effect measurements allowed us to determine the charge carrier density, the upper critical field, the diffusion coefficient, the coherence length and the Ioffe-Regel product $k_Fl $ in the prepared films. The carrier-concentration 
$n\approx 1.7\times 10^{23}$~cm$^{-3}$ does not depend on the thickness of the thin film for t=15, 10 and 5 nm, while sheet resistance $R_{\Box}$ changes considerably from ~110~$\Omega$ for $t=15$~nm through $~200$~$\Omega$ for $t=10$~nm 
to 1100~$\Omega$ for $t=5$~nm. For thickness $t\approx 10$~nm, the 
$k_Fl\approx 2$, indicating that the film is in highly disordered limit. The details of the transport data analysis will be published elsewhere.\cite{Samuely14}

\section{Microwave measurement}
Transmission measurements of the CPW resonators yielded temperature dependencies of the resonant angular frequency $\omega_0$ and the quality factor $Q$, both depending on its complex conductivity.
The CPW resonators were designed to have $\omega_0=2\pi\times 2.5$~GHz and $Q_{ext}\approx4\times10^4$ for conventional superconductor with high thickness. The design of the resonator was verified by a test resonator fabricated out of a thick MoC film (200~nm, Tc=6.7~K). The measured $\omega_0$ and $Q$ completely agreed with the design.
To compare our experimental data  with theory, we calculated the complex impedance of the CPW resonator with known geometry (see Fig.\,\ref{fig:cpw}) using the complex conductivity given by the Mattis-Bardeen theory.\cite{MattisBardeen}

\begin{figure}
		\centering
\includegraphics[width=8cm]{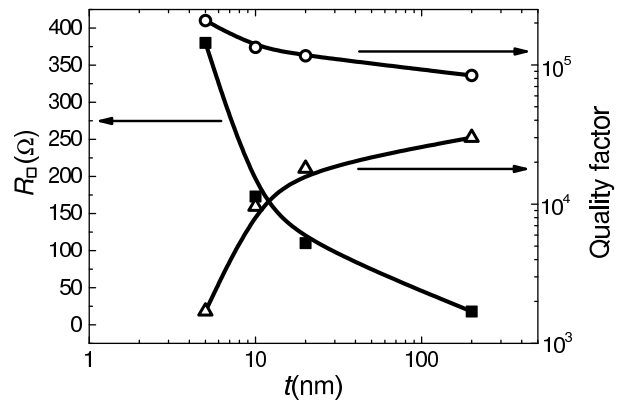}
\caption{The thickness dependence of sheet resistances (squares) at temperatures just above $T_c$ and  internal quality factors (triangles) of MoC coplanar resonators at temperatures $T\ll T_c$. Internal quality factors, limited by dielectric losses,  calculated from the Mattis-Bardeen model for the corresponding $R_{\Box}$ and $t$ (circles) exhibit the opposite trend. The solid lines are
eye-guides.}
\label{fig:samples}
\end{figure}

\begin{figure}
		\centering
	\includegraphics[width=8cm]{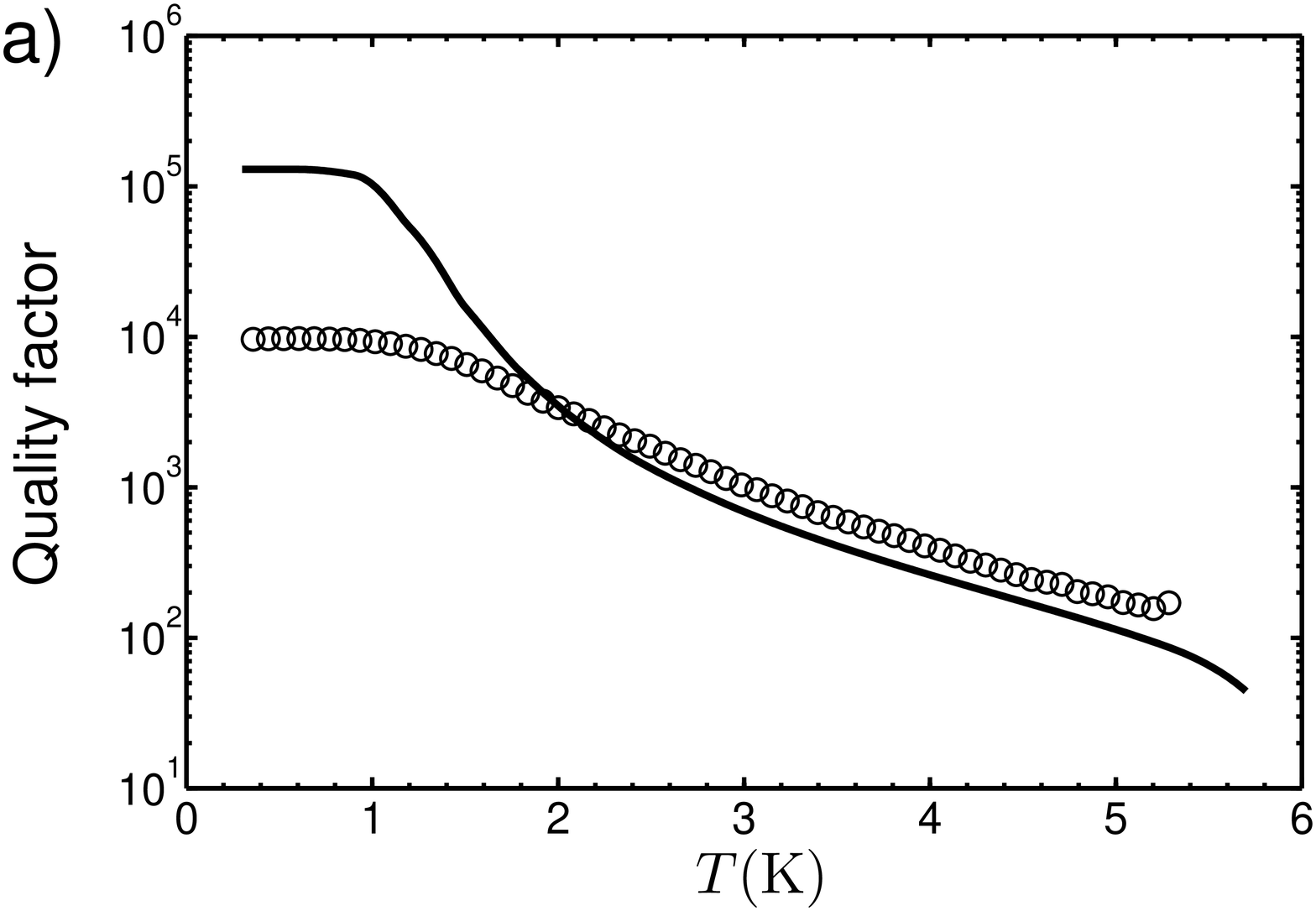}
	\includegraphics[width=8cm]{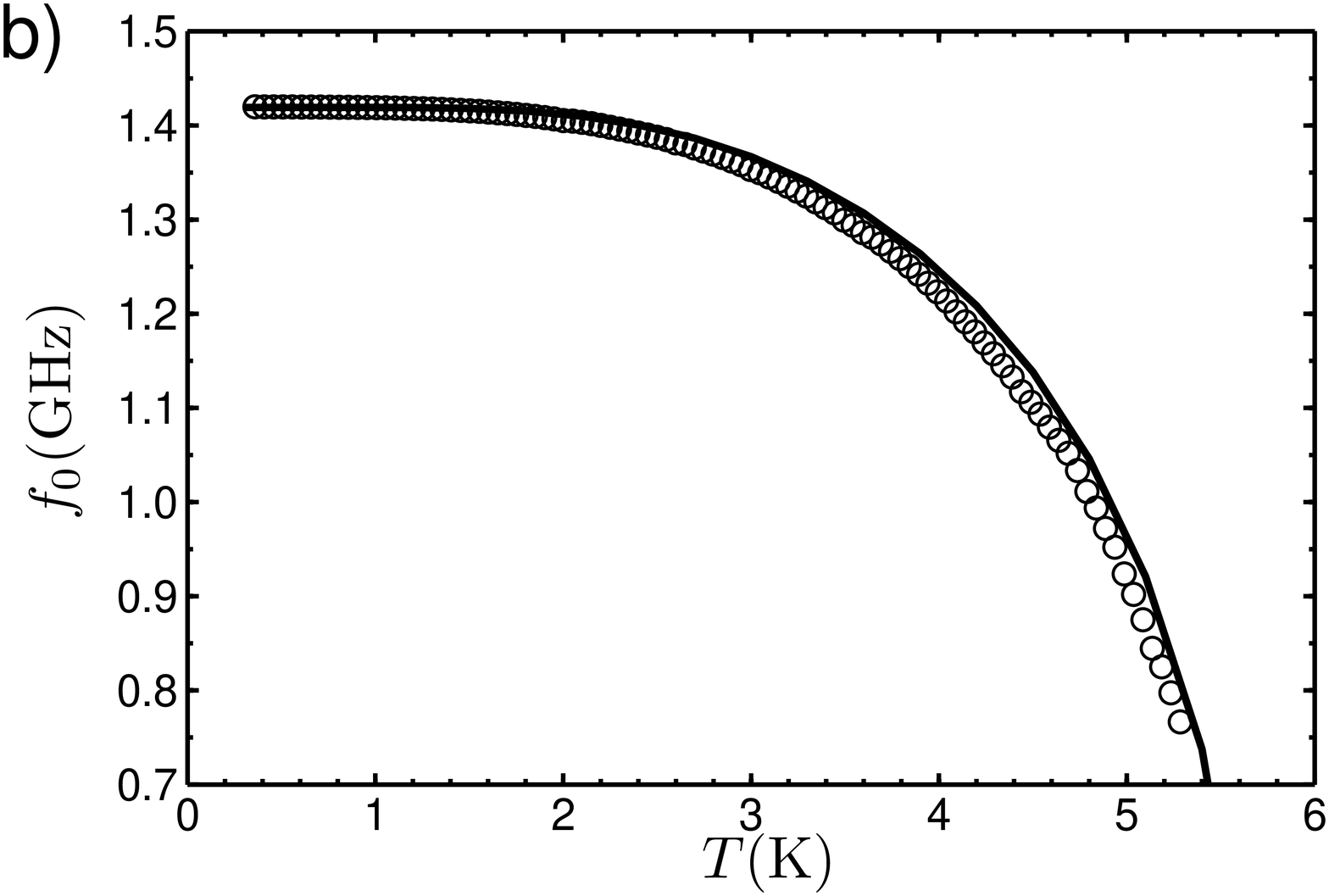}	
\caption{ The temperature dependence of  the quality factor $Q$ (a) and the resonant frequency $f_0$(b) of the 10~nm thin CPW resonator. Circles  are measured data and lines are data calculated from the original Mattis-Bardeen relations ($s~\rightarrow~0$) for parameters $T_c$=5.8~K, $R_\Box=180~\Omega$, $\Delta_0=1.83kT_c$.}
\label{fig:b11a}
\end{figure}

We measured several MoC samples with different thickness (and thus sheet resistance) - their
parameters are presented
in Fig.~\ref{fig:samples}. The most striking feature of our data is that, while
the Mattis-Bardeen model predicts that with an increase of the sheet resistance the quality factor
would increase as well, the experiment reveals an opposite trend: decrease of the measured quality factors with the growth of the sheet resistance. 
Furthermore, the measured quality factor noticeably differ from those predicted by the Mattis-Bardeen model in a wide temperature range, see Fig.~\ref{fig:b11a}a.
At the same time, the measured resonant frequency falls below the theoretical one (Fig.~\ref{fig:b11a}b) only slightly, but systematically.
This deviation was studied in Ref.~\onlinecite{Driessen12} for a narrow temperature range.

It is worth noticing that including mesoscopic fluctuations, which leads to the broadened superconducting density of states,\cite{Feigelman12} does not improve noticeably the agreement between the theory and experiment.

It is clear from Fig. \ref{fig:b11a}a that the loss of a 10~nm CPW resonator at low temperatures is much higher than predicted by the Mattis-Bardeen model, explained below.

\section{Tunneling spectroscopy}
\begin{figure}
\includegraphics[width=8cm]{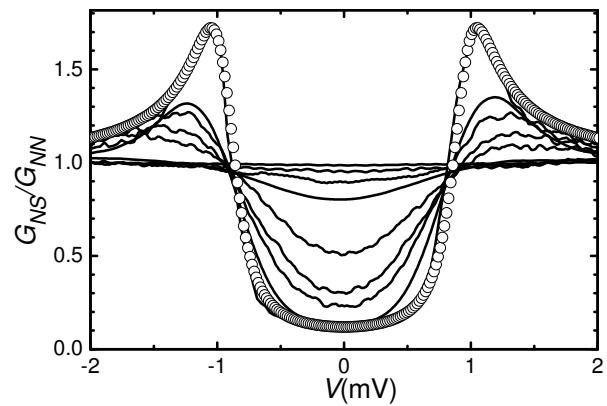}
\caption{The temperature dependence of differential tunneling conductance of the 10~nm MoC thin film measured by a scanning tunneling microscope (solid lines). From the bottom to the top (at zero voltage), the data correspond to temperatures T=0.43, 1, 2, 3, 4, 5, 5.4, 5.6~K. The open circles represent a broadened density of states given by the Dynes formula for $\Gamma=0.12 \Delta_0$, $\Delta_0=1.83kT_c$, $T_c=5.8$~K.}
\label{fig:gamma}
\end{figure}

High losses in disordered superconductors imply a finite density of states at Fermi energy.
Previous studies of disordered superconductors have shown that both microwave measurements and tunneling spectroscopy indicate a broadened superconducting density of states.\cite{Coumou13}
Therefore, we have carried out scanning tunneling spectroscopy  measurements, making use of a low-temperature scanning tunneling microscope (STM). The Fig.~\ref{fig:gamma}
shows the normalized tunneling conductance spectra obtained between
the Au tip and the MoC sample as measured at different
temperatures ranging from 0.43 to 5.8~K.  Each curve was normalized to the spectrum
measured at 5.8~K with the sample  in the normal state in order to exclude the influence of the applied voltage on tunneling barrier and normal density of states of electrodes. Therefore each of these
normalized differential conductance versus voltage spectra reflects the
superconducting density of states (SDOS) of MoC sample, smeared by $2k_BT$ in energy at the respective temperature. Consequently, at the low temperature limit ($k_BT\ll\Delta$), the differential conductance measures the SDOS directly.
As evidenced by the 0.43 K curve, the measured SDOS differs from the BCS SDOS: it reveals significant quasiparticle density of states at the Fermi level and broadened coherence peaks at the gap edges.
In our case the best agreement with the experimental data is obtained for the empiric Dynes formula (\ref{eq:Dynes}) with parameters 
$\Gamma=0.12\Delta_0$, $\Delta_0=1.83kT_c$, $T_c=5.8$~K. Here we want to emphasized that the experimentally obtained superconducting density of states is broadened, but spatially uniform in lateral directions. We have taken STS spectra along 200 nm line and the coefficient of variation of the normalized tunneling conductance at zero voltage and superconducting energy gap is about 0.05 and 0.025, respectively.  There is no characteristic length-scale of variations and they can be attributed to a noise and instabilities of the measuring system during long scanning measurements.

\section{Modification of the Mattis-Bardeen theory}
Building on the extended BCS theory\cite{BCS} Mattis and Bardeen (MB) derived frequency-dependent complex conductivity.\cite{MattisBardeen}
To do so, they formally introduced an infinitesimal
scattering parameter, $s=2\pi/\tau$, which was set to zero at the end of calculations.
We may conjecture that in disordered superconductors the finite value of $s$ may acquire the physical meaning of inverse quasiparticle lifetime and use
the corresponding expressions derived in Ref.~\onlinecite{Belitz91}, where both, phonon and Coulomb contributions to the quasiparticle lifetime were taken into account.
Keeping a finite value of $s$, as a phenomenological inverse inelastic
quasiparticle lifetime, one can derive the modified formulas for the ratio of the superconducting complex conductivity $\sigma_s$ to the normal conductivity $\sigma_n$ as

\begin{equation} \label{eq:sigma}
\frac{\sigma_s}{\sigma_n}=1-\frac{1}{\hbar\omega}\int_{\Delta}^\infty [1-2f(E)]g(E,\omega)dE,
\end{equation}

where $f(E)$ is the Fermi-Dirac distribution function and the propagator $g$ is defined as

\begin{eqnarray} \label{eq:g}
g(E,\omega)&=&g_+(E,\omega)-\mathrm{csgn}(E-(\hbar\omega-is))g_-(E,\omega) \nonumber \\
g_\pm(E,\omega)&=&\frac{E}{\sqrt{(E^2-\Delta^2)}}
\frac{E\pm\hbar(\omega-is)}{\sqrt{(E\pm\hbar(\omega-is))^2-\Delta^2}}\nonumber \\
&+& \frac{\Delta}{\sqrt{(E^2-\Delta^2)}}
\frac{\Delta}{\sqrt{(E\pm\hbar(\omega-is))^2-\Delta^2}}.
\end{eqnarray}

\begin{figure}
\includegraphics[width=8cm]{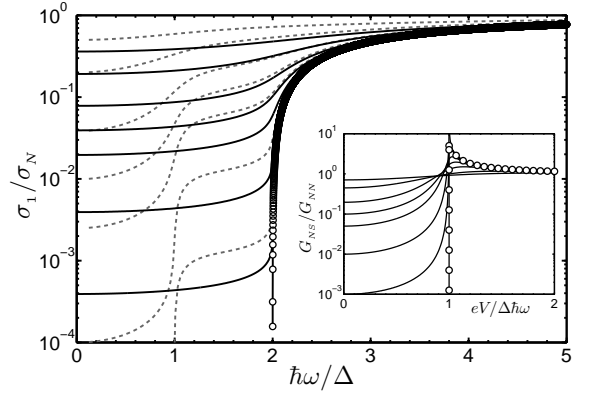}	
\caption{Frequency dependence of the real part of the superconducting complex conductivity $\sigma_1$ for the finite inelastic scattering parameters. Circles are the original Mattis-Bardeen (scattering parameter $s=0$). Solid and dashed lines are theoretical curves corresponding, from bottom to top, to the finite values $\hbar s/\Delta, \Gamma/\Delta=0.001, 0.01, 0.05, 0.1, 0.2, 0.5, 1$ for  Nam and MB model with finite scattering, respectively; \\ \textit{inset:} The normalized tunneling conductance of the normal metal-insulator-superconductor tunnel junction for the same values of $\Gamma=\hbar s$.}
\label{fig:sig1_inset}
\end{figure}

\begin{figure}
		\centering
\includegraphics[width=8cm]{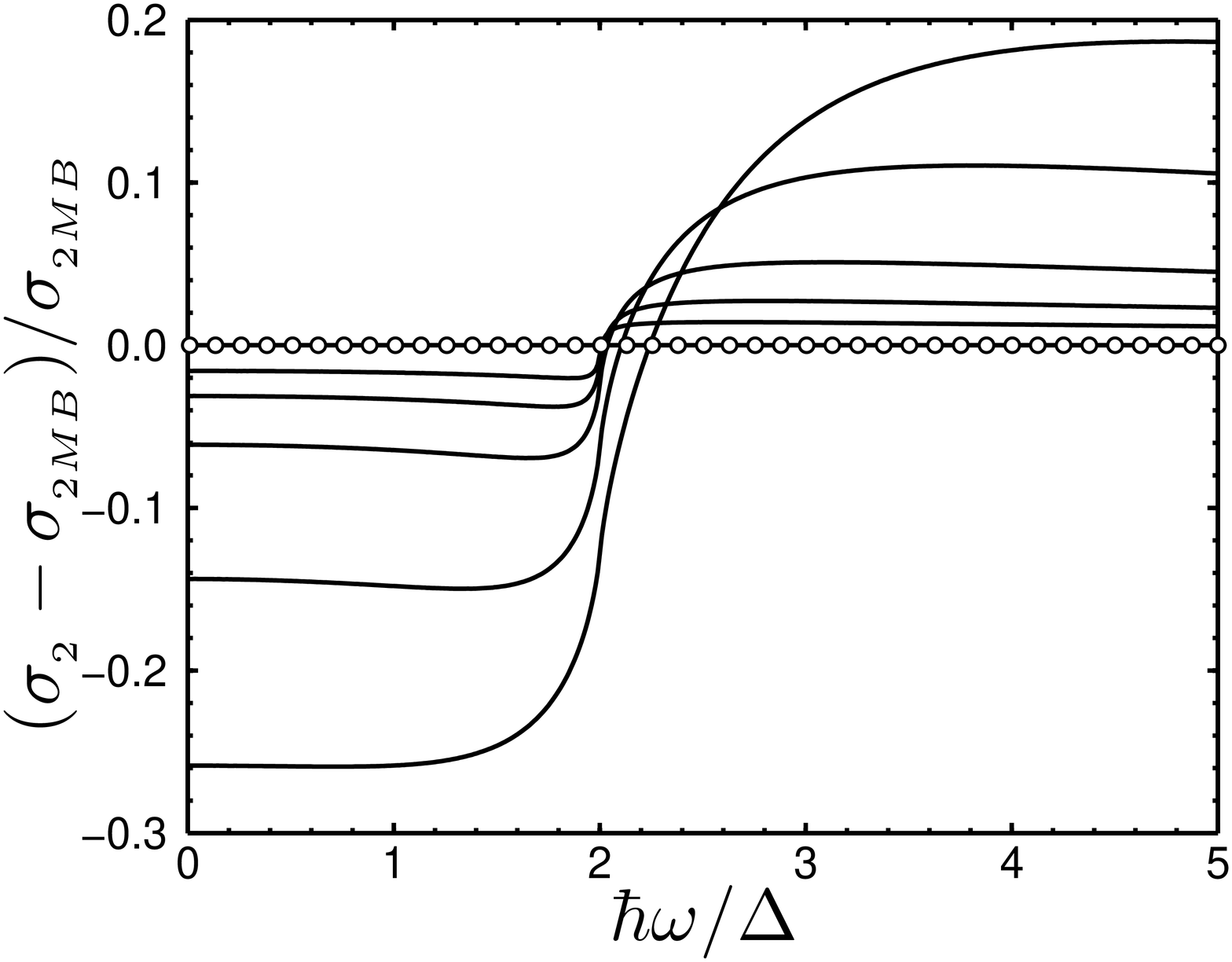}	
\caption{Relative deviation of the imaginary part of the superconducting complex conductivity $\sigma_2$ from the original Mattis-Bardeen theory (circles) for the same values of parameter $\hbar s$ as in Fig \ref{fig:sig1_inset}. Increasing of the parameter $s$ decreases the magnitude of the $\sigma_2$ for frequencies below $2\Delta$.}
\label{fig:sig2fr}
\end{figure}

Here $E$
is the quasiparticle energy, $\Delta$ is the superconducting energy gap, and $\mathrm{csgn}(E-(\omega-is))$ is the complex signum function, defined as $\mathrm{csgn}(z)\equiv 1,-1,\mathrm{sgn}(\Im(z))$ for $\Re(z)>0$, $\Re(z)<0$, and $\Re(z)=0$, respectively.
Inspecting Eq.\,(\ref{eq:g}), one sees that the first term of $g_\pm(E,\omega)$ is a product
of the standard BCS quasiparticle density of states and a similar factor, but with
broadened energy states.  The broadening can be viewed
as the result of Coulomb and/or phonon interactions. 
It is to be noted, however, that the derived propagators $g_\pm(E,\omega)$ include the same broadened density of states as described by
the empiric Dynes formula used in tunneling and point contact spectroscopy\cite{Dynes84,Grajcar94}

\begin{equation}
N(E)={\Re} \left\{\text{sgn}(E)\frac{E+i\Gamma}{\sqrt{(E+i\Gamma)^2-\Delta^2}}\right\},
\label{eq:Dynes}
\end{equation}

with the inelastic scattering parameter $\Gamma=\hbar s$.
The superconducting complex conductivity and the tunneling conductance
for finite inelastic scattering parameters are shown in Figure~\ref{fig:sig1_inset},\ref{fig:sig2fr}.
Although MB model with finite scattering provides results 
consistent with microwave measurements, the results contain 
a product of BCS and broadened superconducting density of states, which seems to be nonphysical. Nevertheless, one can include the broadened density of states
to Nam model\cite{Nam67a} elaborated for superconductors containing magnetic impurities. Nam generalized the Mattis-Bardeen model for arbitrary complex functions $n(E)$ and $p(E)$ whose real parts correspond to the densities of states and Cooper pairs, respectively. For Dynes broadened density of states,
the complex functions $n(E)$, $p(E)$ can be defined as 
\begin{eqnarray} \label{eq:np}
n(E)&\equiv &\text{sgn}(E)\frac{E+i\Gamma}{\sqrt{(E+i\Gamma)^2-\Delta^2}}\nonumber \\
p(E)&\equiv &\text{sgn}(E)\frac{\Delta}{\sqrt{(E+i\Gamma)^2-\Delta^2}}
\end{eqnarray}    
and the normalized superconducting complex conductivity reads
\begin{eqnarray} \label{eq:sigmaNam}
\frac{\sigma_1}{\sigma_n}=\frac{1}{\hbar\omega}\int_{-\infty}^\infty &[f(E)&-f(E+\hbar\omega)][\Re(n(E))\Re(n(E+\hbar\omega))\nonumber \\ &+&\Re(p(E))\Re(p(E+\hbar\omega))]dE \nonumber \\ \nonumber \\
\frac{\sigma_2}{\sigma_n}=\frac{1}{\hbar\omega}\int_{-\infty}^\infty &[1-&2f(E+\hbar\omega)][\Im(n(E))\Re(n(E+\hbar\omega))\nonumber \\ &+&\Im(p(E))\Re(p(E+\hbar\omega))]dE
\end{eqnarray}
The square roots are taken to mean the principal square root with the real part greater than or equal to zero.

The real parts of the complex conductivity calculated for Nam model and the modified Mattis-Bardeen model with finite scattering are compared with standard Mattis-Bardeen model in Fig.~\ref{fig:sig1_inset}. Both models result in an increase of the real part of the complex conductivity in comparison to the standard MB model. At frequencies $\omega>2\Delta$ this increase is almost indistinguishable, whereas at frequencies $\omega<2\Delta$ they differ significantly. The difference is the most noticeable at $\omega=\Delta$, where the Nam model exhibit a peculiarity. The resonant frequency of our resonator is much lower than the energy gap $\omega\ll\Delta$, therefore 
we cannot test the peculiarity at $\omega\approx\Delta$ directly. The THz spectroscopy performed recently on NbN samples \cite{Sherman15} could detect such peculiarity but measurement temperatures are too high to resolve it.

\begin{figure}
\centering
\includegraphics[width=8cm]{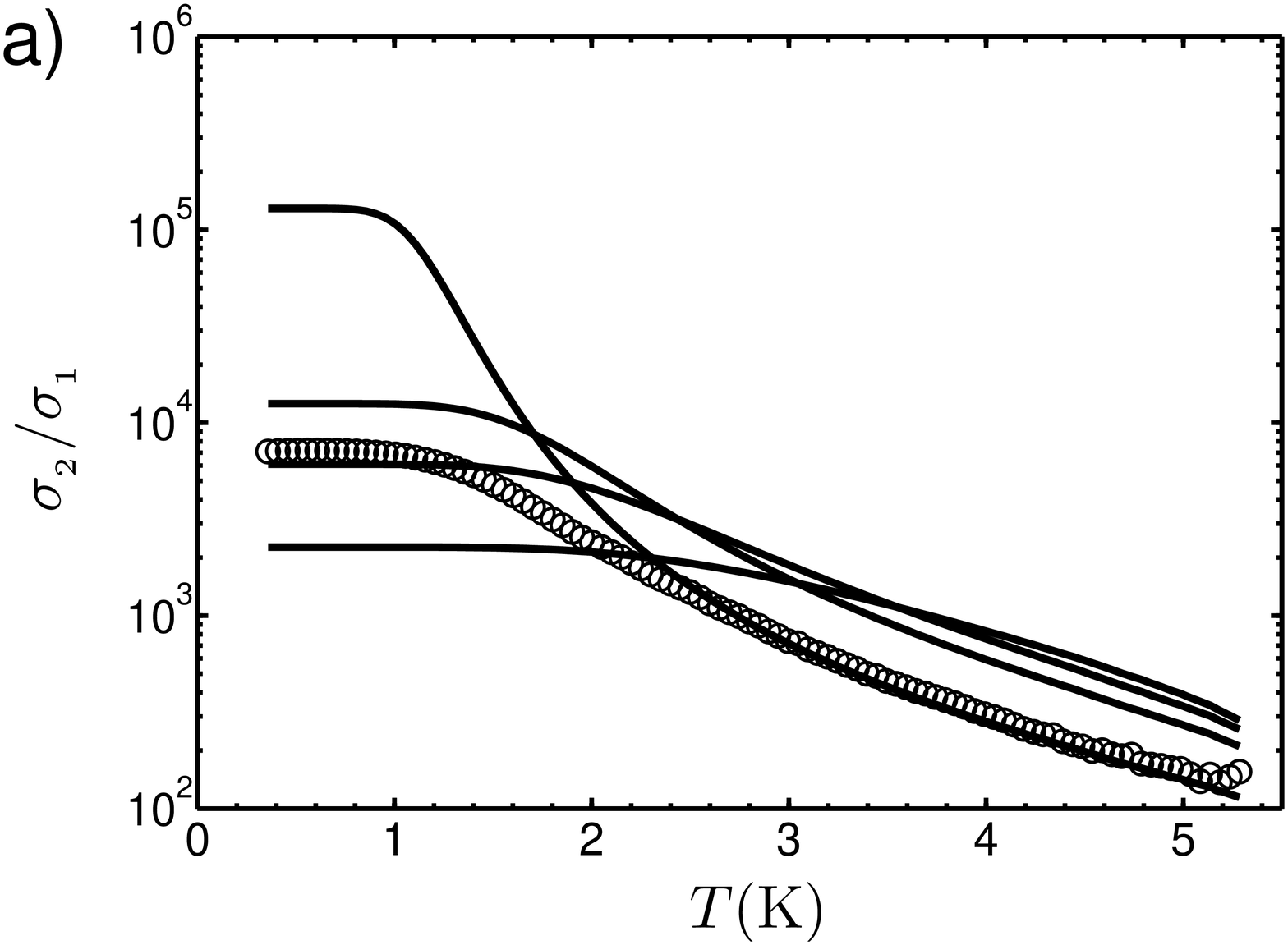}
\includegraphics[width=8cm]{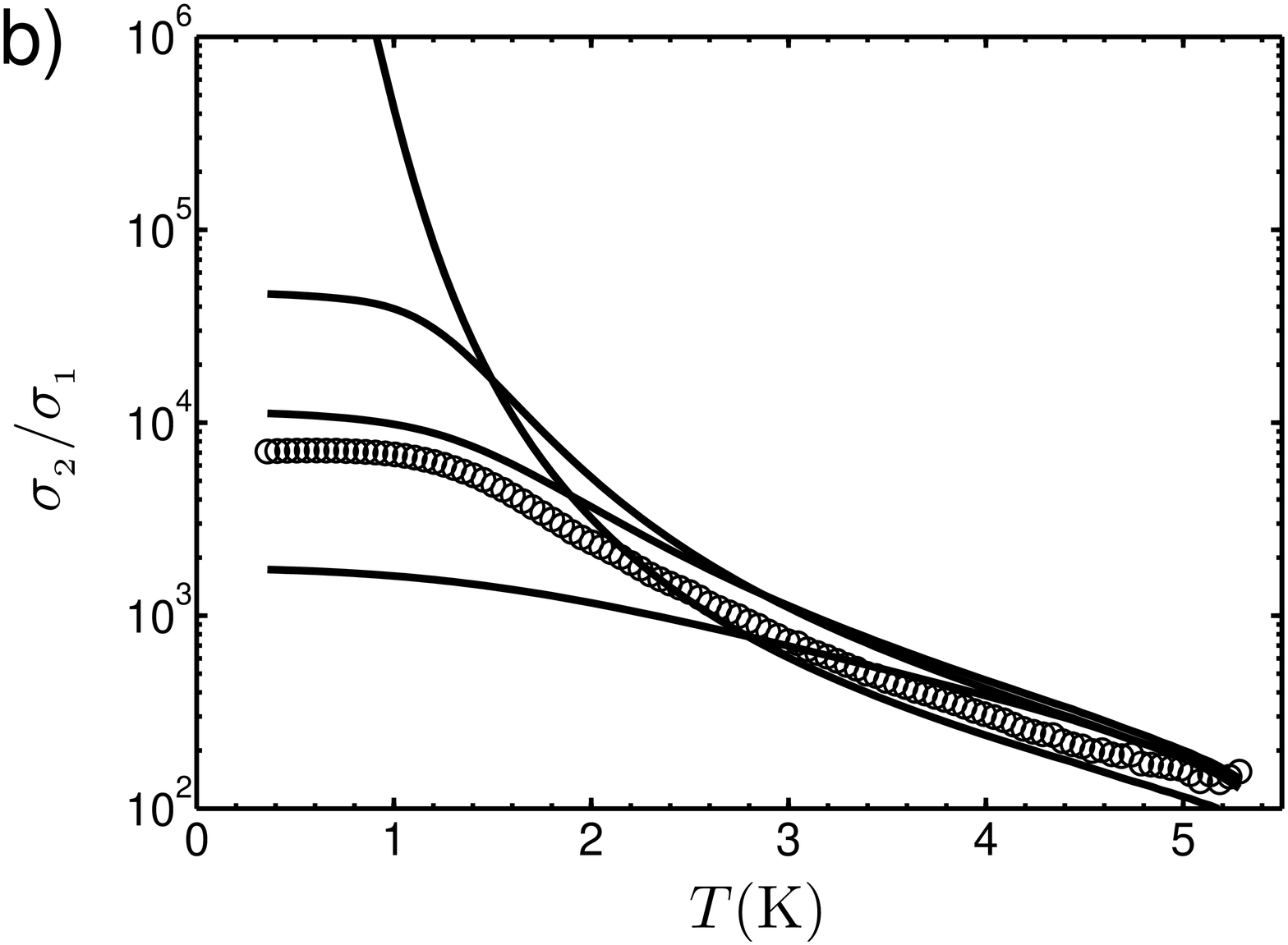}
\caption{Measured ratio $\sigma_2/\sigma_1$ (open circles) for the 10~nm thin MoC CPW resonator compared with the ratio calculated from the Mattis-Bardeen (upper) relations with finite inelastic scattering and Nam model (lower) for various values of the parameter $s$ and $\Gamma$ respectively. At low temperatures (below 1~K) the lines from top to bottom correspond to $\hbar s/\Delta_0, \Gamma/\Delta_0=0.01, 0.1, 0.2, 0.5$.
\label{fig:nam_vs_matbar}}
\end{figure}

In order to avoid problems with normalization and geometrical factor inaccuracy it is convenient to compare theoretical and experimental results via ratio $\sigma_2/\sigma_1$. Since the geometrical factors of the resonator are cancelled out, this ratio can be expressed as a function of the resonant frequency and the quality factor of the CPW resonator 	
\begin{equation}
\frac{\sigma_2}{\sigma_1}=Q\left(1-\left(\frac{\omega_0}{\omega_g}\right)^2\right)\,.
\end{equation}
Here $\omega_g$ is the resonant frequency of the resonator in the normal state of a lossless metal.
In Fig.\,\ref{fig:nam_vs_matbar} we compare the experimental data with the  $\sigma_2/\sigma_1$ temperature dependence calculated for different values of the parameter $s$ and $\Gamma$ corresponding to Mattis-Bardeen and Nam model respectively.

At high temperatures $T>T_c/2$, small values of the scattering parameter provide
a fair agreement between the experimental data and the prediction by the standard MB theory, i.e. this theory works well.
However, at very low temperatures larger value of scattering parameter is required to fit the measured data, and it is not possible to find any intermediate value of 
$s$ or $\Gamma$ to obtain a reasonable agreement with both tunneling and microwave experiments for the
complete temperature range.

\section{Two-channel models}
\begin{figure}
	\centering
\includegraphics[width=8cm]{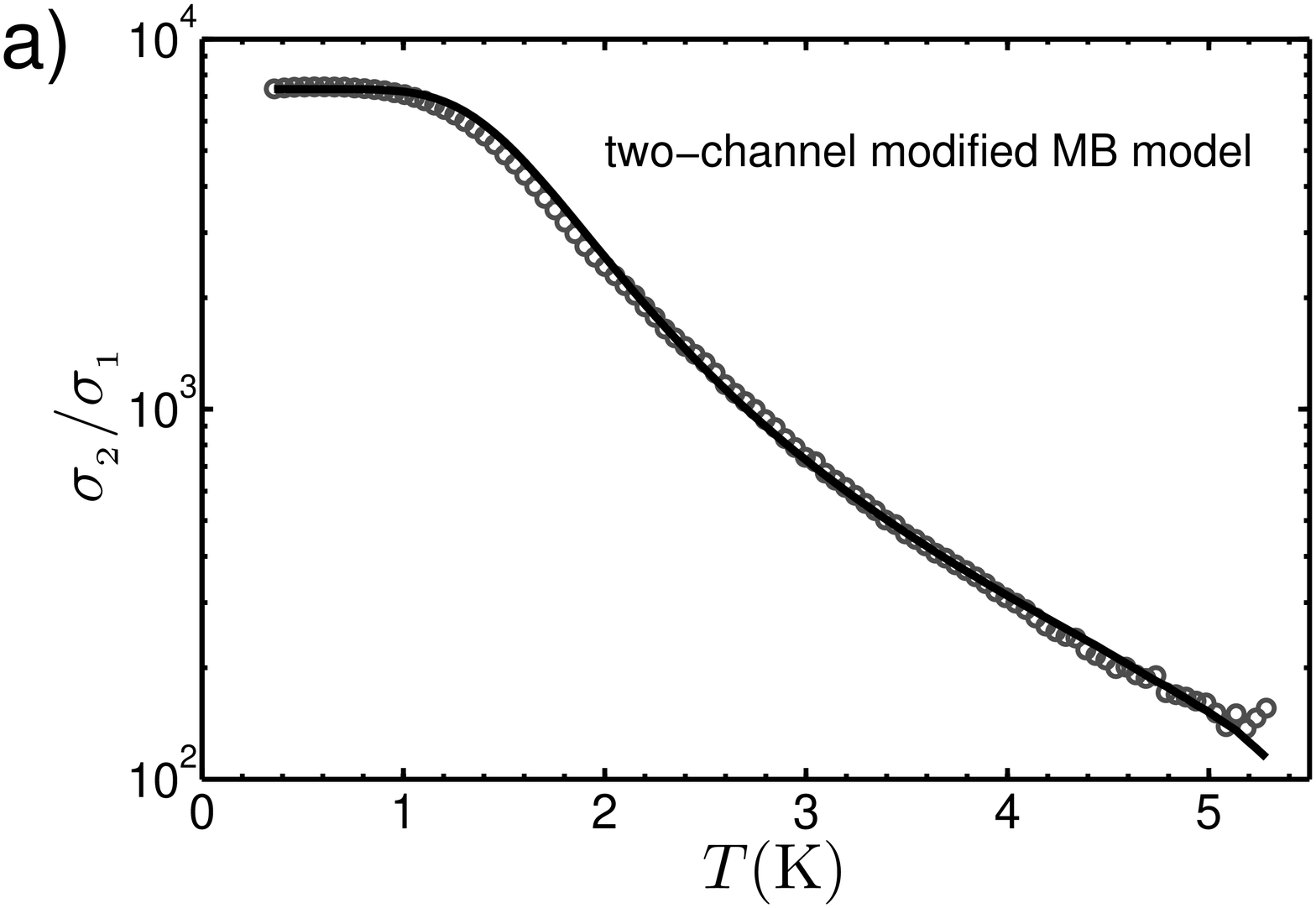}
\includegraphics[width=8cm]{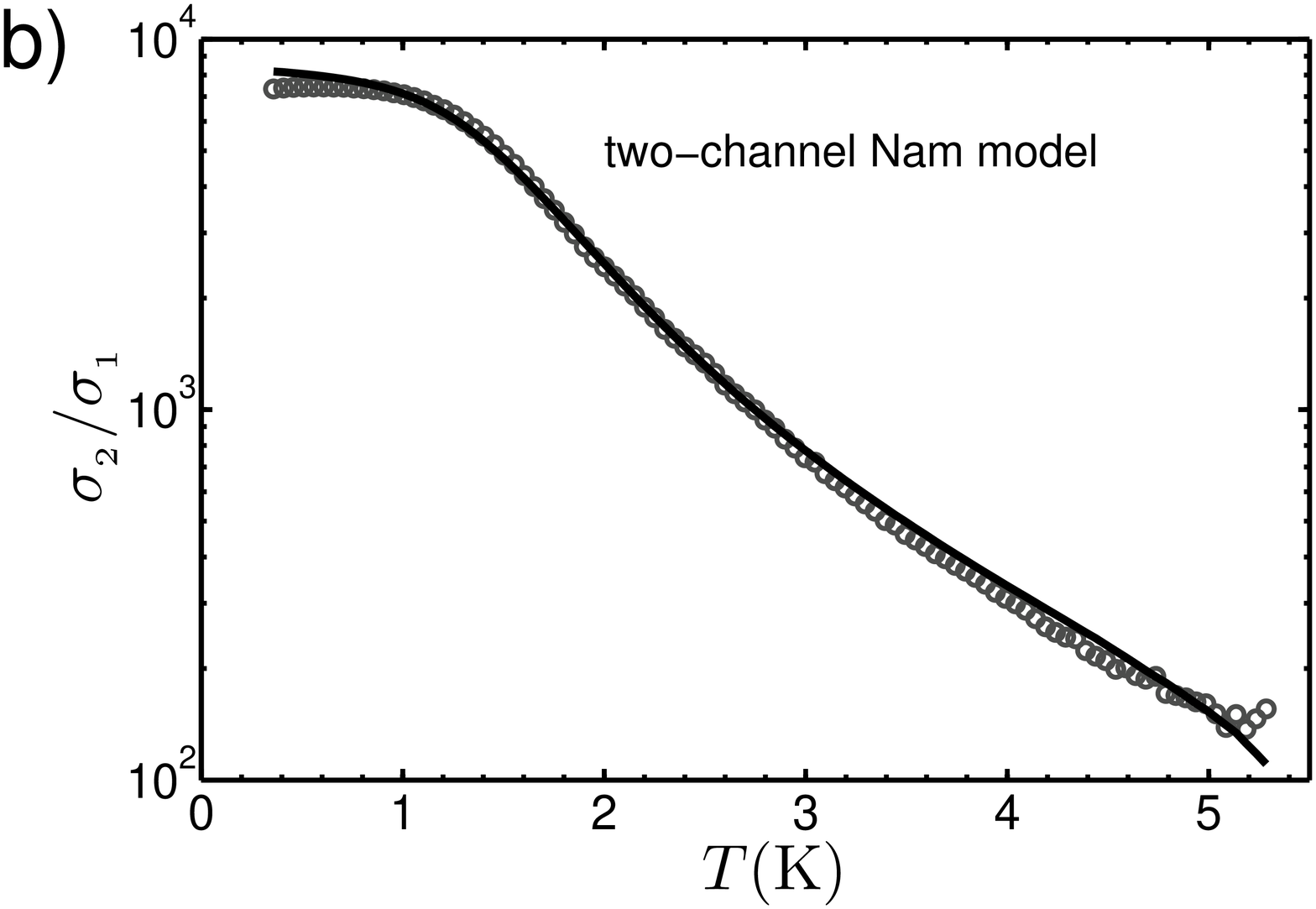}

\caption{The temperature dependence of the ratio $\sigma_2/\sigma_1$ of the 10~nm thin CPW resonator.
Circles are measured data fitted by the two channels model with (a) modified Mattis-Bardeen relations (solid line) for parameters $\kappa=0.34$, $\Delta_0=1.83kT_c$,
$T_c=5.7$~K, $\hbar s_s=0.24\Delta_0$, $s_b=0$, (b) Nam relations (solid line) for parameters $\kappa=0.10$, $\Delta_0=1.83kT_c$, $T_c=5.7$~K, $\Gamma=0.25\Delta_0$}
\label{fig:fit}
\end{figure}

The MB model with finite scattering parameter $s$ provides a hint how to solve the problem. In order to obtain a self-consistent picture, one should include also a ``channel" corresponding to the photon scattering from the BCS to the BCS density of states which corresponds to the standard MB model with $s\rightarrow0$. 
Hence let us adopt a two channel model in which the total complex conductivity is a weighted sum of two contributions $\sigma=(1-\kappa)\sigma_{b}+\kappa\sigma_{s}$,
where the $\sigma_{s}$ corresponds to the channel with enhanced scattering $s_s$ taken as fitting parameter while the one without scattering (bulk one) $s_b$ is taken to be zero and the parameter $\kappa\in\langle 0,1\rangle$ is the filling factor. 
Surprisingly, the modified MB model with
a weighted sum of two contributions gives excellent agreement with experimental results, see Fig.~\ref{fig:fit}a.
Here we can conclude that in order to fit the complex conductivity and tunneling conductivity of disordered superconductors for a similar set of parameters, one should apply the two channel model - this conclusion was also reached by D. Sherman et al. in Ref.~\onlinecite{Sherman15}.
The same procedure was applied to Nam model and the results are shown in Fig. \ref{fig:fit}b. The results are satisfying for temperatures above $T_c/2$, but the qualitative agreement between theory and experiment for small temperatures is much worse than those for the modified MB model.

\begin{figure}
	\centering
\includegraphics[width=8cm]{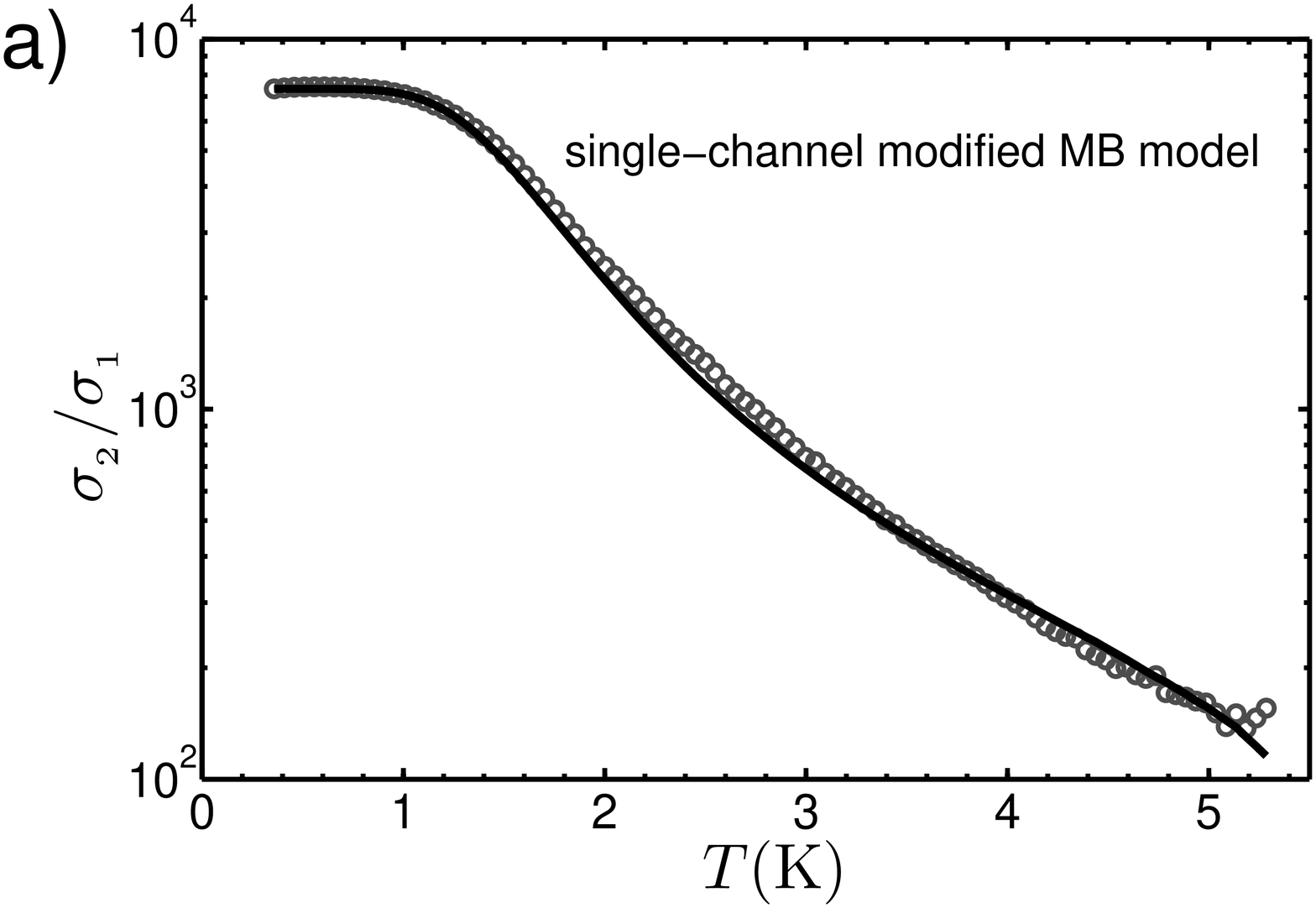}
\includegraphics[width=8cm]{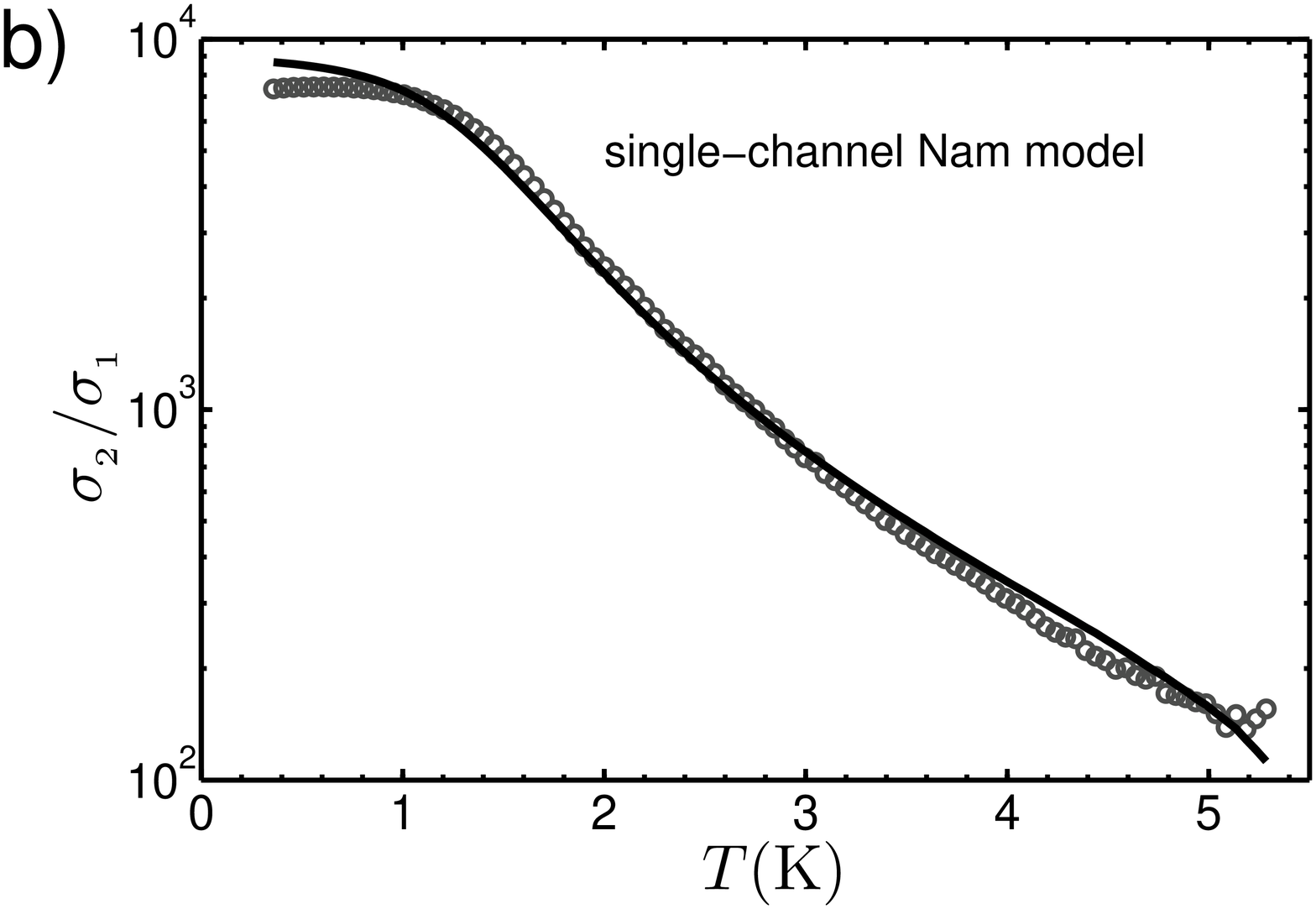}

\caption{Same experimental data as in Fig. \ref{fig:fit} (circles) fitted by the one channel model with (a) modified Mattis-Bardeen relations (solid line) for parameters $\Delta_0=1.5kT_c$, $T_c=5.7$~K, $\hbar s=0.13\Delta_0$, (b) Nam relations (solid line) for parameters $\Delta_0=1.7kT_c$, $T_c=5.7$~K, $\Gamma=0.23\Delta_0$}
\label{fig:fit2}
\end{figure}

Remarkably, if we adopt an argument of Sherman et al.\cite{Sherman14} that the metallic tip of the scanning tunneling microscope can screen Coulomb interactions which can, in turn, increase the measured energy gap, it is not necessary to fit the complex conductivity for the same values of the energy gap as measured by STM. Hence, the $\Delta_0$ can be taken as a fitting parameter and a smaller value than the one obtained from the STM is expected. The best fit of the complex conductivity was obtained  for $\kappa \rightarrow 0$, which is in fact a single channel model. The obtain suppressed gap is $\Delta_0\approx 1.5 kT_c$, which is much smaller than the $\Delta_0\approx 1.83 kT_c$ measured by STM, whereas the scattering parameter $\Gamma\approx 0.13 \Delta_0$, is very close to the value obtained by STM (see Fig.~\ref{fig:fit2}a).
Interestingly, for Nam theory the same approach with single channel and suppressed $\Delta_0$ does not change the results in comparison with the two channel model (see Fig. \ref{fig:fit2}b).
For both single channel models the ratio $\Delta/kT_c$ is below BCS universal value 1.76.
Nevertheless, in both cases, single or two-channel, considerable inelastic scattering is 
required which fully justifies the concept of finite quasiparticle lifetime in disordered superconductors. 
Surprisingly, Nam models do not fit the experimental data as good as the modified MB ones. Therefore, it would be interesting to check both models experimentally at $\omega\approx\Delta/\hbar$, where they exhibit the most remarkable difference. THz spectroscopy can provide such experimental data.

\begin{figure}
	\centering
\includegraphics[width=8cm]{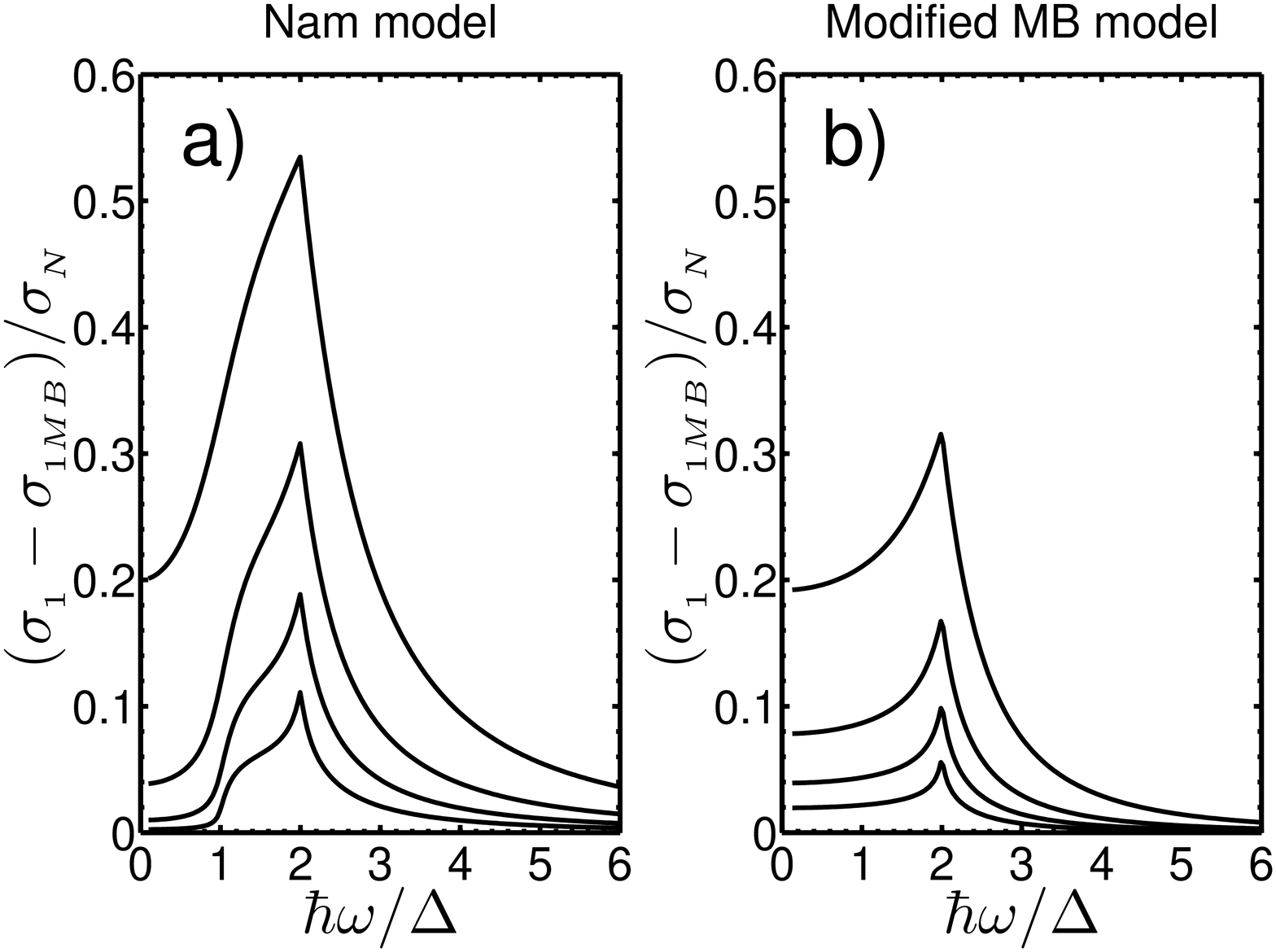}
\includegraphics[width=8cm]{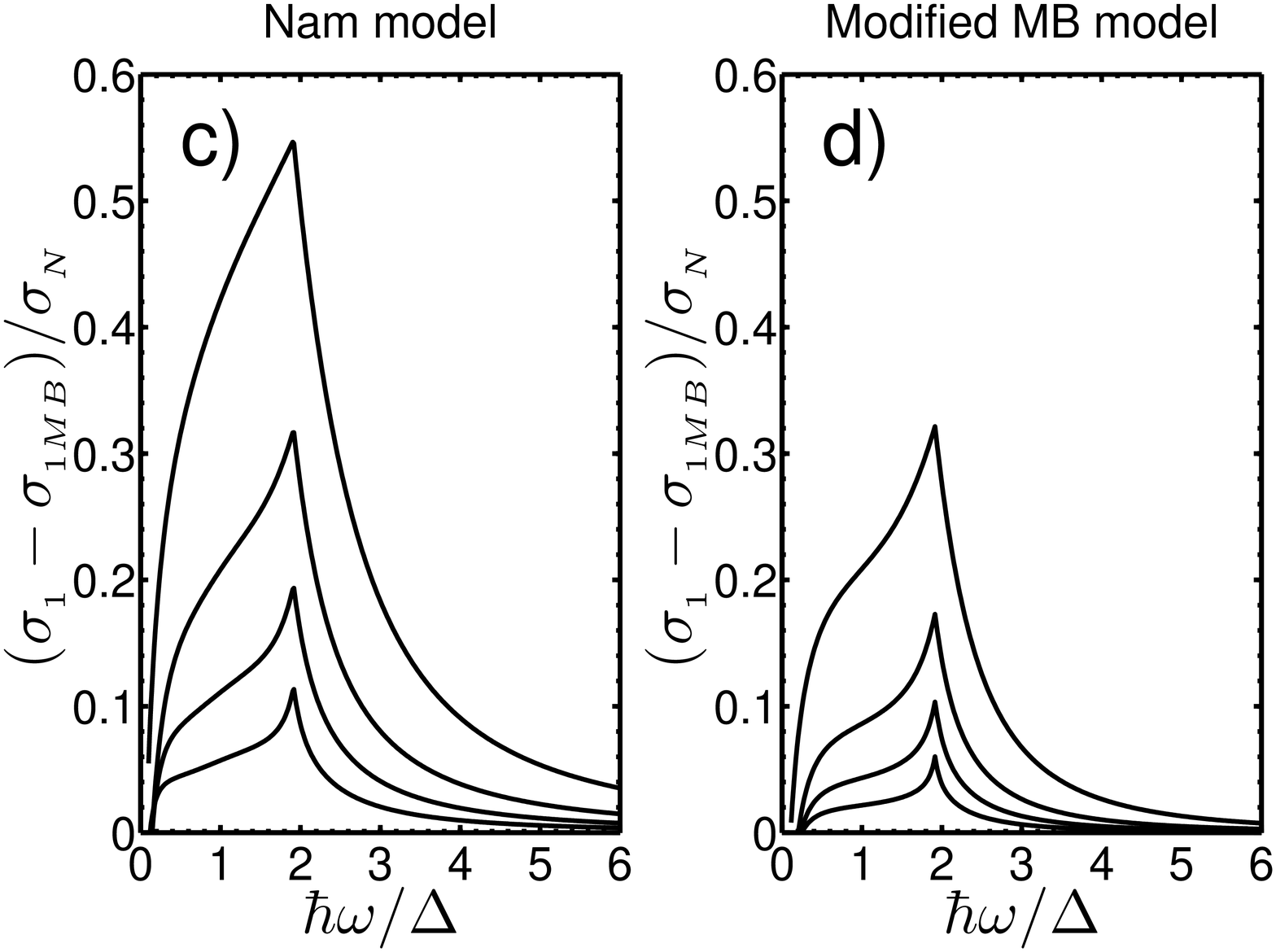}
\caption{Normalized deviation of the real part of the complex conductivity $\sigma_1$ to the standard Mattis-Bardeen counterpart ($s=0$) for Nam (a, c) and modified MB model (b, d) for temperatures $T=0$ (a, b) and $T=0.4 T_c$ (c, d). From bottom to top the curves in each graph correspond to inelastic scattering parameters: $\Gamma/\Delta_0, \hbar s/\Delta_0 =$ 0.05, 0.1, 0.2, 0.5.}
\label{fig:thzdif}
\end{figure}

In Ref.~\onlinecite{Sherman15}, the THz spectroscopy reveals  a deviation of the measured real part of the complex conductivity $\sigma_1$ from the standard Mattis-Bardeen counterpart. The difference, which authors ascribe to a contribution of broken symmetry in disordered superconductors, is compared with Higgs model.\cite{Swanson14} However the deviations presented in Fig.~3 in 
Ref.(~\onlinecite{Sherman15}) can be even better described by models with broadened density of states. For example the position of the peaks clearly changes with the energy gap of the samples and there is no cutoff frequency  at which the deviation of the measured real part of the complex conductivity from the standard MB model $\sigma_1-\sigma_{1MB}$ saturates to zero. Both features are present in Fig.~\ref{fig:thzdif}, where the deviation of the theoretical curves $\sigma_1(\omega)$ from standard MB model ($s=0$) are shown for various inelastic scattering parameters. 
These results show that one should take into account the broadened density od states in MB model and more exotic models should be compared to the modified MB model.

Our two-channel model is phenomenological and does not address a problem of microscopic origin of the second channel with enhanced inelastic scattering. 
Nevertheless, the Dynes empiric formula for broadened density of states,
introduced in 1984 for disordered superconductors,\cite{Dynes84} is present in MB model and Mattis-Bardeen had it already in 1958. Such broadened density of states was observed in many superconducting system as disordered superconductors,\cite{Dynes84} high T$_c$ superconductors,\cite{Grajcar94} MgB$_2$,\cite{Karapetrov01} iron based superconductors,\cite{Szabo09} etc.. It seems that, for some paths, the scattering parameter $s$ is not renormalized to zero and Mattis-Bardeen model with finite scattering rate $s$ is a good first approximation. The scattering is probably caused by two-level systems located very homogenously at the MoC-sapphire interface which are present even if special precautions are made.\cite{Bruno15} This hypothesis is corroborated by our scanning tunneling microscopy/spectroscopy which reveal the same inelastic scattering parameter $\Gamma$ even for atomically flat surfaces with no adsorbed impurity.       

\section{conclusions}
In conclusion,  the two channel model well describes both microwave and tunneling conductance measurements in a wide temperature range - from 300~mK up to almost $T_c$. The enhanced scattering channel dominates at low temperatures, which is consistent with the low temperature losses in high quality CPW resonators made of conventional superconductors,\cite{Macha10,Bruno15} where the quality factor is limited by the interface scattering. However, if STM provides overestimated value of the superconducting energy gap, the results can even be described by the one channel model with enhanced scattering. In disordered superconducting films such scattering leads to the suppression of electron diffusion and, accordingly, to enhancement of the Coulomb interaction.\cite{Finkel87,Finkel94} This is consistent with the rapid decrease of the quality factors for thinner superconducting films as shown in Fig.~\ref{fig:samples}.

The main features shown on MoC disordered superconductors apply also for other disordered superconductors, such as NbN or TiN.\cite{Coumou13,Sherman15} For example, the deviation of the real part of complex conductivity $\sigma_1$ from the standard Matttis-Bardeen one ($s=0$), measured in Ref.~\onlinecite{Sherman15}, can be well explained by finite inelastic scattering as shown in Fig.~\ref{fig:thzdif}, providing even better agreement with experiment. Thus, the original Mattis-Bardeen model with finite inelastic scattering or the Nam model are able to describe microwave and tunneling experiments in disordered superconductors, while more advanced models are failing. Our results providing a simple expression for the complex 
conductivity call for further theoretical, as well as experimental research aimed at clarifying what does the simple MB theory with finite inelastic scattering captures that is lost in more advanced machineries.\newline

\textit{Acknowledgments}: This work was supported by the European Community's Seventh Framework Programme (FP7/2007-2013) under Grant No. 270843 (iQIT), by the MP-1201 COST Action, by the Slovak Research and Development Agency under the contract DO7RP–0032–11, APVV-0515-10, APVV-0808-12, APVV-14-0605, and by the U.S. Department of Energy, Office of Science, Materials Sciences and Engineering Division. P.Sz. and P.S. acknowledge the Slovak Research and Development Agency Contract No. APVV-0036-11, and VEGA 1/0409/15. EI acknowledges a partial support by the Russian Ministry of Science and education, Contract No. 8.337.2014/K.

\end{document}